\documentclass[usenatbib]{mn2e}
\usepackage{amsfonts}
\usepackage{amsmath}
\usepackage{graphicx}
\usepackage{natbib}

\title[Shapley and the Sloan Great Wall]
      {How unusual are the Shapley Supercluster and the Sloan Great Wall?}

\author[R. K. Sheth \& A. Diaferio]
       {Ravi K. Sheth$^{1,2}$\thanks{E-mail: shethrk@physics.upenn.edu, diaferio@ph.unito.it} \& Antonaldo Diaferio$^{3,4,5}$\\
 $^1$Department of Physics \& Astronomy, University of Pennsylvania, 
 209 S. 33rd Street, Philadelphia, PA 19104, USA\\
 $^2$The Abdus Salam International Center for Theoretical Physics, 
      Strada Costiera 11, 34151 Trieste, Italy\\
 $^3$Dipartimento di Fisica Generale ``Amedeo Avogadro'', Universit\`a degli Studi di Torino, Via P. Giuria 1, I-10125, Torino, Italy\\
 $^4$Istituto Nazionale di Fisica Nucleare (INFN), Sezione di Torino, Via P. Giuria 1, I-10125, Torino, Italy\\
 $^5$Harvard-Smithsonian Center for Astrophysics, 60 Garden Street, Cambridge, MA 02138, USA}

\newcommand{\bm}[1]{{\mbox{\boldmath $#1$}}}

\begin{document}
\pagerange{\pageref{firstpage}--\pageref{lastpage}}

\maketitle

\label{firstpage}

\begin{abstract}
We show that extreme value statistics are useful for studying the
largest structures in the Universe by using them to assess the
significance of two of the most dramatic structures in the local
Universe -- the Shapley supercluster and the Sloan Great Wall.
If we assume that the Shapley concentration 
(volume $\approx 1.2\times 10^5 h^{-3}$Mpc$^3$)
evolved from an overdense 
region in the initial Gaussian fluctuation field, 
with currently popular choices for the background cosmological
model and the shape and amplitude $\sigma_8$ of the initial power 
spectrum, we estimate that the total mass of the system is within
20 percent of $1.8\times 10^{16}h^{-1}\,M_\odot$.  
Extreme value statistics show that the existence of this massive 
concentration is not unexpected if the initial fluctuation field was 
Gaussian, provided there are no other similar objects within a sphere 
of radius $200h^{-1}$Mpc centred on our Galaxy.
However, a similar analysis of the Sloan Great Wall, a more distant 
($z\sim 0.08$) and extended concentration of structures
(volume $\approx 7.2\times 10^5 h^{-3}$Mpc$^3$) suggests that it is 
more unusual.  We estimate its total mass to be within 20 percent of
 $1.2\times 10^{17} h^{-1}M_\odot$
and we find that even if it is the densest such object of its volume
within $z=0.2$, its existence is difficult to reconcile with the
assumption of Gaussian initial conditions if $\sigma_8$ was less
than 0.9.  This tension can be alleviated if this structure is the 
densest within the Hubble volume.  
Finally, we show how extreme value statistics can be used to address 
the question of how likely it is that an object like the Shapley 
Supercluster exists in the same volume which contains the Sloan 
Great Wall, finding, again, that Shapley is not particularly unusual. 
Since it is straightforward to incorporate other models of the 
initial fluctuation field into our formalism, we expect our approach 
will allow observations of the largest structures -- clusters, 
superclusters and voids -- to provide relevant constraints on the 
nature of the primordial fluctuation field.
\end{abstract}


\section{Introduction}
Since its discovery \citep{discov} the Shapley Supercluster 
has been the object of considerable interest because it potentially 
contributes significantly to the velocity field in the local Universe 
\citep[e.g.][]{scaramella,somak}
and because the existence of extremely massive objects such as Shapley 
constrains the amplitude of the initial fluctuation field, and possibly 
the hypothesis that this field was Gaussian.

Recent studies suggest that the Shapley Supercluster contains a few 
times $10^{16}h^{-1}\,M_\odot$, is overdense by a factor of order 2, 
and is receding from us at about 15,000 km~s$^{-1}$.  
These conclusions are based on studies of the motions of galaxies 
\citep{quintana,r2000,proust,ragone06} and estimates of the masses 
of X-ray clusters in this region \citep{reiprich02,fse05}.  
In addition, the fact that this region is over-abundant in rich 
clusters also allows an estimate of its mass \citep{ml08}, 
not all of which may actually be bound to the system 
\citep{dunner,future}.  
Whereas the other methods are observationally grounded, the mass 
estimate from this last method (i.e. from the over-abundance of 
rich clusters) follows from the assumption that the initial 
fluctuation field was Gaussian.  Here, we refine this estimate of 
the total mass of Shapley and compare it with the answer to the question:  
What is the probability distribution of the mass of the most massive 
object, having the volume of Shapley, if it formed from Gaussian 
initial conditions?  
We use extreme value statistics to address this question.  
Although we do not explore this here, we note that our methods are 
easily extended to incorporate non-Gaussian initial conditions.  

Section~\ref{Shapley} summarizes a number of properties of the 
Shapley supercluster.  
Sections~\ref{NmDelta} and~\ref{extremes} describe our methods
based on the excursion set approach and extreme value statistics, 
and what they imply for objects like Shapley, for which accurate 
estimates of the masses of the constituent clusters are available.  
Section~\ref{sdss} shows how to extend these approaches to study the 
Sloan Great Wall \citep{gott05}, for which accurate mass estimates of 
the components are not available.  
This requires combining a halo model \citep[e.g.,][]{review} analysis 
of the galaxy population with a catalog of groups identified in this 
distribution.  For the SDSS, we use the clustering and group analyses 
of Zehavi et al. (2005) and Berlind et al. (2006), respectively.  

A final section summarizes our results, shows how extreme value 
statistics can be used to answer the question of how unusual it 
is that an object like the Shapley Supercluster exists in the same 
volume which contains the Sloan Great Wall, and discusses how 
our methods allow observations of the largest structures -- clusters, 
superclusters and voids -- to place interesting constraints on the 
nature of the initial fluctuation field.  
Where necessary we assume a flat $\Lambda$CDM model with 
$(\Omega_0,\Omega_b,h,\sigma_8) = (0.27,0.046,0.72,0.8)$, but we also 
explore other choices of $\sigma_8$.  

\section{The Shapley Supercluster}\label{Shapley}
\begin{figure}
 %\begin{center}
 \hspace{-7cm}
 \includegraphics[scale=0.9,angle=90]{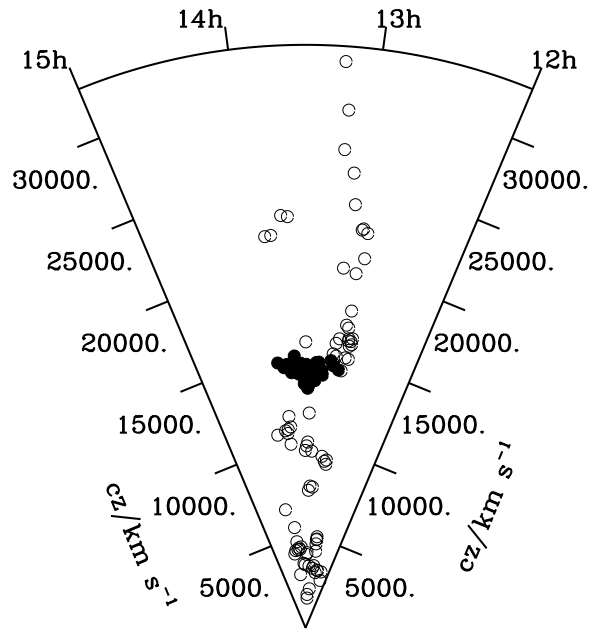}
 \vspace{-5cm}
 \caption{Systems of galaxies of the \citep{ragone06} sample in 
     redshift space.  Solid dots are the clusters belonging to the 
     Shapley supecluster according to a percolation analysis with 
     percolation length $8 h^{-1}$~Mpc.}
 \label{fig:ragone}
 %\end{center}
\end{figure}

The largest redshift survey which includes the Shapley supercluster 
suggests that it contains 8632 galaxies \citep{proust}.  These have 
been grouped into 122 systems of galaxies with 4 or more members 
\citep{ragone06}.  We run a percolation algorithm on this catalog 
to identify the largest supercluster in this region.  To do so, we 
neglect the peculiar velocity of the clusters: i.e., each cluster 
is assigned coordinates
 $x_1 = r\cos\delta\cos\alpha$, $x_2=r\cos\delta\sin\alpha$ and 
 $x_3=r\sin\delta$, where 
 $(\alpha,\delta)$ are its celestial coordinates and $r=cz/H_0$.
Figure~\ref{fig:ragone} shows the pie diagram of these systems. 
Solid dots show the 40 systems belonging to the Shapley Supercluster 
when we use a linking length of $8 h^{-1}$~Mpc.  According to the 
virial masses computed by Ragone et al., 15 of these 40 clusters 
have masses larger than $10^{14} h^{-1}$~M$_\odot$.  Summing the masses 
of these 40 clusters yields $5.42 \times 10^{15} h^{-1}$~M$_\odot$.  
The total mass is expected to be considerably larger than this, 
because lower mass groups and galaxies are expected to contribute 
significantly to the total.  Ragone et al. (2006) use mock catalogs, 
based on the VLS simulation of Yoshida et al. (2001), to account for 
this missing mass, and conclude that the total mass of Shapley is 
likely to be about $1.6\times 10^{16}h^{-1}M_\odot$.  

To quantify the shape of the Shapley supercluster, we compute 
the eigenvalues of the inertia tensor 
\begin{equation}
 I_{ij} = {\sum_k m_k x_{ki} x_{kj}\over \sum_k m_k} 
        \quad {\rm where}\quad i,j=1,2,3
\end{equation}
where $m_k$ is the mass of each cluster, the coordinates $x$ 
are centered on A3558, and the sum is only over the cluster members. 
We find the three eigenvalues $8.30$, $5.48$, and $2.73 h^{-1}$~Mpc. 
If we neglect the fact that the 40 cluster members have masses in the 
range $[0.008, 6.717]\times 10^{14}h^{-1}$~M$_\odot$,
and set $m_k=1$ for all $k$'s, the eigenvalues of the tensor of 
inertia are $7.69$, $6.02$, and $3.42 h^{-1}$~Mpc; i.e., they are 
not substantially different from the previous values.

As a check, we have also applied our percolation analysis to an 
X-ray survey of this region, which shows 41 extended sources 
\citep{fse05}.  
A link length of $8h^{-1}$~Mpc links 8 clusters, and returns a total 
mass in X-ray clusters of 
 $1.65 \times 10^{15} h^{-1}$~M$_\odot$, 
where we estimated the mass of each cluster as follows:
\begin{equation}
 {M_{200}\over h_{50}^{-1} M_\odot} = 
  \left( L_{\rm bol}\over 10^{A+40} h_{50}^{-2} {\rm erg}\;
                          {\rm s}^{-1}\right)^{1/\alpha}
\end{equation}
where $A=-22.1\pm 1.3$ and $\alpha=1.807\pm 0.084$
 \citep{reiprich02}.\footnote{This differs slightly 
from Mu\~noz \& Loeb (2008), who assume that
 $M_{\rm 200}\propto L^{1/1.6}$.}
With this recipe, only 5 out of the 8 members have masses larger than 
$10^{14} h^{-1}$~M$_\odot$.  It is reassuring that these numbers are 
smaller than those of Ragone et al. (2006), because this sample of 
X-ray clusters with known redshifts is clearly incomplete \citep{fse05}.  
Therefore, in what follows, we use the cluster catalogue
from Ragone et al. (2006), rather than from the X-ray data.

\section{The excursion set approach}\label{NmDelta}
The previous section suggests that the total mass of the Shapley 
supercluster is at least $5\times 10^{15} h^{-1}M_\odot$.  In this 
section, we make a rather different estimate of the total mass.  
According to Ragone et al. (2006), the inner $31h^{-1}$Mpc of Shapley centered 
on A3558 contains 58 galaxy systems: 19 of these have mass greater 
than $10^{14}h^{-1}M_\odot$.  For such high masses, it is reasonable 
to equate each cluster with a single halo.  
Integrating the halo mass function \citep{st99} from this lower 
limit to infinity shows that the expected number in randomly 
placed spheres of this radius is only $2.67$.  This number depends 
on $\sigma_8$:  reducing $\sigma_8$ to 0.7 changes the expected count 
to 1.77; increasing to 0.9 makes the count 3.5.  
Neither of these numbers is close to that observed.  

However, if Shapley is an overdense region, then the relevant 
comparison is not with the expected counts in a region of average 
density, but one which is overdense \citep{ml08}.  
In theories of structure formation from Gaussian initial conditions, 
massive halos are expected to be more abundant in dense regions, 
and the mix of halos is expected to also be different.  In dense 
regions, the halo mass function is expected to be top-heavy 
\citep{fewd88, mw96, st02}, so this is an immediate signal that Shapley 
must be overdense in dark matter \citep{ml08}.  Measurements in 
the SDSS indicate that the halo mass function in regions which are 
overdense in galaxies is indeed top-heavy \citep{sscs06,as07}, so 
it is interesting to ask if this effect is sufficient to explain the 
existence of a region like Shapley.  

\begin{figure}
 \begin{center}
 \includegraphics[scale=0.433]{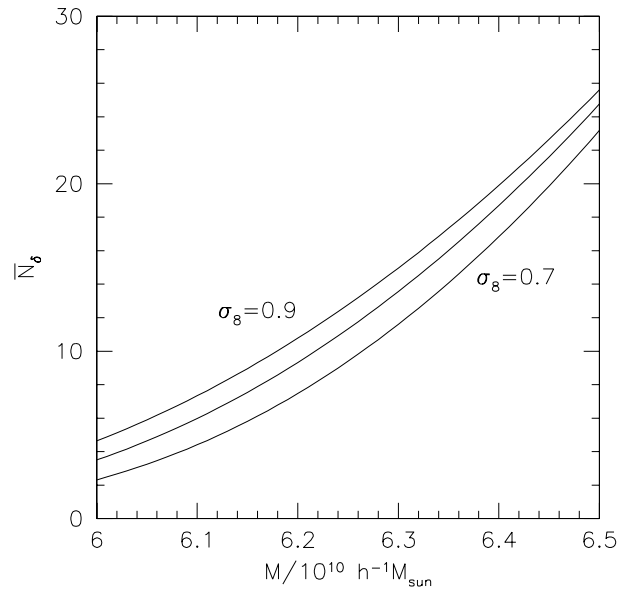}
 \caption{Expected number of clusters with masses greater than 
          $10^{14}h^{-1}M_\odot$ as a function of the total mass 
          of the supercluster.  The expected number increases as 
          $\sigma_8$ increases.  }
 \label{fig:meanN}
 \end{center}
\end{figure}

To make this estimate, we will make the crude assumption that 
Shapley is spherical, despite the fact that it is not, as we
have shown in the previous section. However, by considering the 
most massive 19 clusters within a distance of $31h^{-1}$Mpc from A3558, 
rather than the system identified with the percolation analysis, we
expect to make this assumption more reasonable.  We will return to 
the issue of triaxiality in the final Discussion section.

Let $\bar N_\delta$ denote the mean number of halos with mass above 
threshold in a region which has volume $V$ and contains mass $M$ 
(so the mass overdensity is $1+\delta = M/\bar\rho V$):
\begin{equation}
 \bar N_\delta = \int_{M_{\rm min}}^M {\rm d}m \, N(m,\delta_c|M,V).
 \label{Ndelta}
\end{equation}
This number increases as $M$ increases; the precise dependence can 
be computed following arguments in Sheth \& Tormen (2002), which 
build on the work of Mo \& White (1996), and are within the 
framework of the excursion set approach \citep{lc93,bcek91}.\footnote{Note 
that the procedure followed by Mu{\~n}oz \& Loeb (2008) for 
estimating $\bar N_\delta$ will yield large-scale halo bias factors 
which are the same as those of Mo \& White (1996); these are known 
to be inaccurate \citep{st99}.  
Our procedure produces bias factors which are in substantially 
better agreement with simulations.}  
This approach requires an estimate of the relation between the 
overdensity in linear theory, $\delta_{\rm L}$ and the actual 
nonlinear overdensity $1+\delta$.  We use the spherical model to 
do this:
\begin{equation}
 1+\delta \approx 
 \left(1 - \frac{\delta_{\rm L}}{\delta_{\rm sc}} \right)^{-\delta_{\rm sc}},
 \label{scapprox}
\end{equation}
where $\delta_{\rm sc}\approx 1.675$.  

Let $p(M|V)$ denote the probability that a randomly placed cell 
of size $V$ contains mass $M$.  If we assume that halo counts in 
cells of mass $M$ follow a Poisson distribution with mean 
$\bar N_\delta$ \cite[see][for why this is only accurate for 
large cells]{sl99}, then the probability that a cell of size $V$, 
in which there are $N$ clusters, contains mass $M$ is 
\begin{equation}
 p(M|N,V) = \frac{p(N|M,V)\,p(M|V)}{p(N|V)},
 \label{pMNV}
\end{equation}
where 
\begin{equation}
 p(N|V) \equiv \int dM\,p(N|M,V)\,p(M|V),
\end{equation}
and the Poisson assumption means 
\begin{equation}
 p(N|M,V) \equiv \frac{\bar N_\delta^N}{N!}\, \exp(-\bar N_\delta).
 % p(N|M,V) \equiv \frac{\bar N_\delta^N}{N!}\, \exp(-\bar N_\delta).
 % if we make a Binomial distribution with mean
 % <N> = Np and <N(N-1)> = Np(N-1)p = <N>^2 - <N>p 
 % N!/n!/(N-n)! p^n (1-p)^(N-n)
 % <n> = Np (N-1)!/(n-1)!/(N-n)! p^(n-1) (1-p)^(N-n)
 % <n(n-1)> = Np (N-1)p
 % 
 \label{poisNMV}
\end{equation}
To proceed, we require a model for the probability $p(M|V)$ that 
a randomly placed cell of size $V$ contains mass $M$.  Now, $p(M|V)$ 
can be estimated using the same excursion set framework as is used 
in the calculation of $\bar N_\delta$ \citep{rks98}.  Alternatively, 
on large scales, it could also be estimated using perturbation 
theory \citep{ptreview}.  On these large scales, these two approaches 
are in good agreement:  the shape of $p(M|V)$ which results is 
reasonably well approximated by a Lognormal \citep{ls08}:
\begin{equation}
 p(M|V)\,{\rm d}M \approx \frac{\exp(-y^2/2\sigma_{\rm L}^2)}
                           {\sigma_{\rm L}\,\sqrt{2\pi}}\,\frac{{\rm d}M}{M},
 \label{pLN}
\end{equation}
where $y = \ln(1+\delta) + \sigma^2_{\rm L}/2$, and 
$\sigma_{\rm L}^2$ is the variance in linear theory on scale $V$.  
For $\sigma_8=(0.7,0.8,0.9)$ and $V=(4\pi/3) 31^3h^{-3}$Mpc$^3$ our linear power spectrum yields 
$\sigma_{\rm L}=(0.23,0.26,0.29)$.

\begin{figure}
 %\begin{center}
 \includegraphics[scale=0.433]{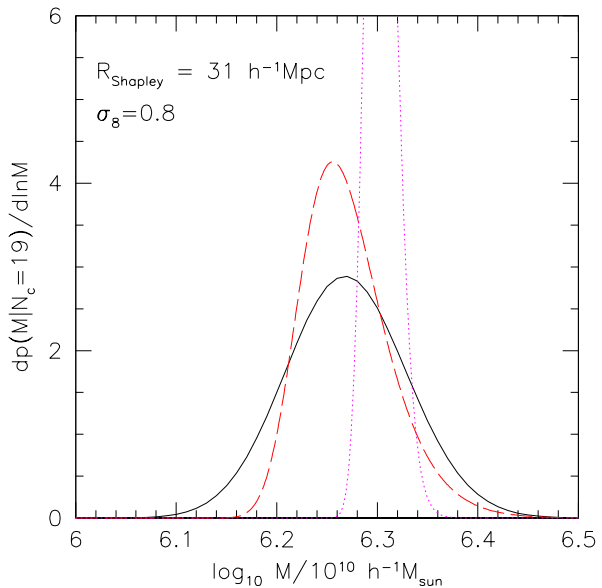}
 \caption{Comparison of the excursion set estimate of the mass of the 
          Shapley supercluster (solid) with the expected mass of the 
          densest of $N = (200/31)^3$ and the sixth densest of 
          $N = (575/31)^3$ randomly placed cells having the same volume 
          as Shapley (dashed and dotted), when $\sigma_8=0.8$.}
 \label{fig:shapley8}
 %\end{center}
\end{figure}

\begin{figure}
 \begin{center}
 \includegraphics[scale=0.433]{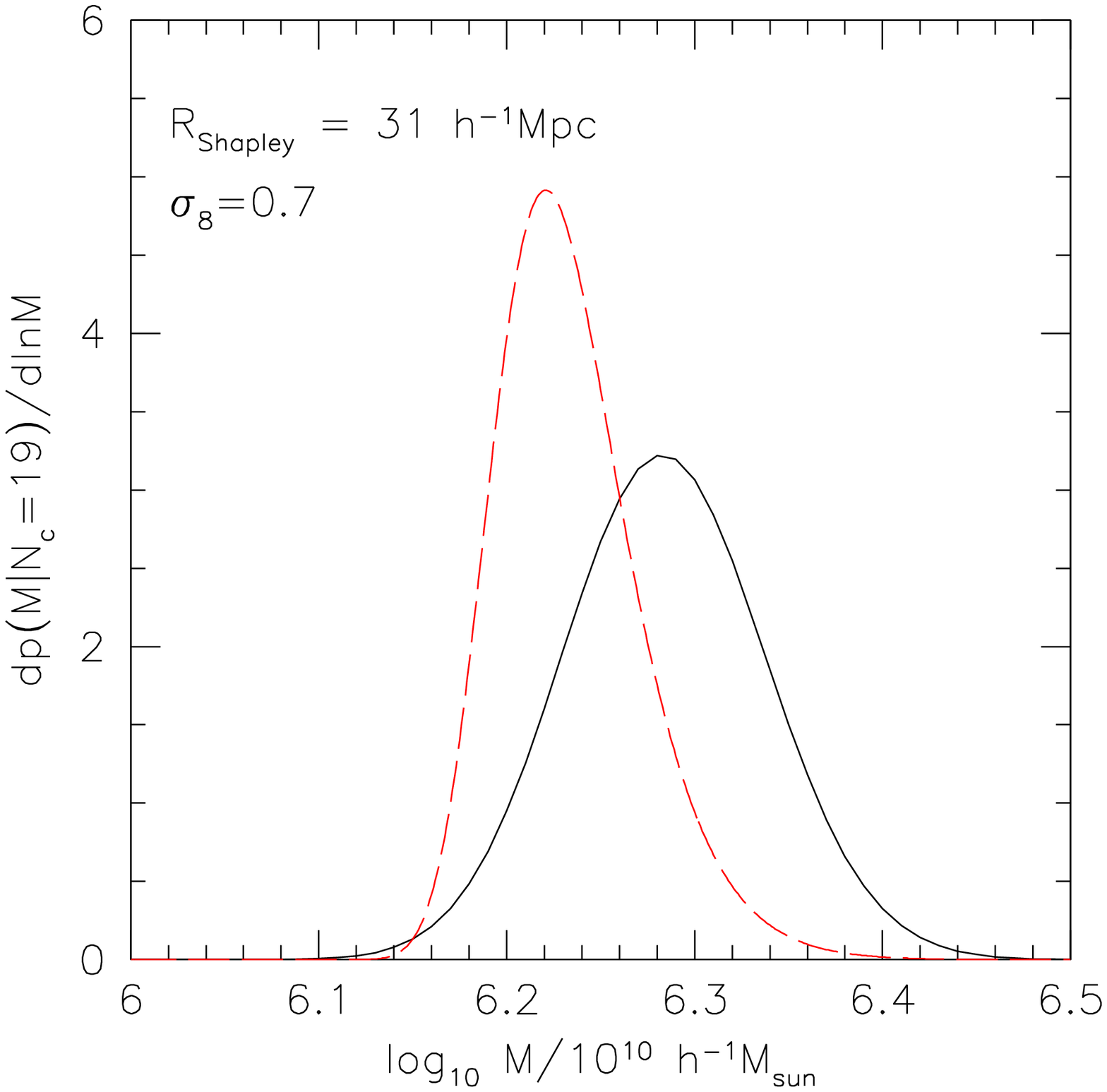}
 \includegraphics[scale=0.433]{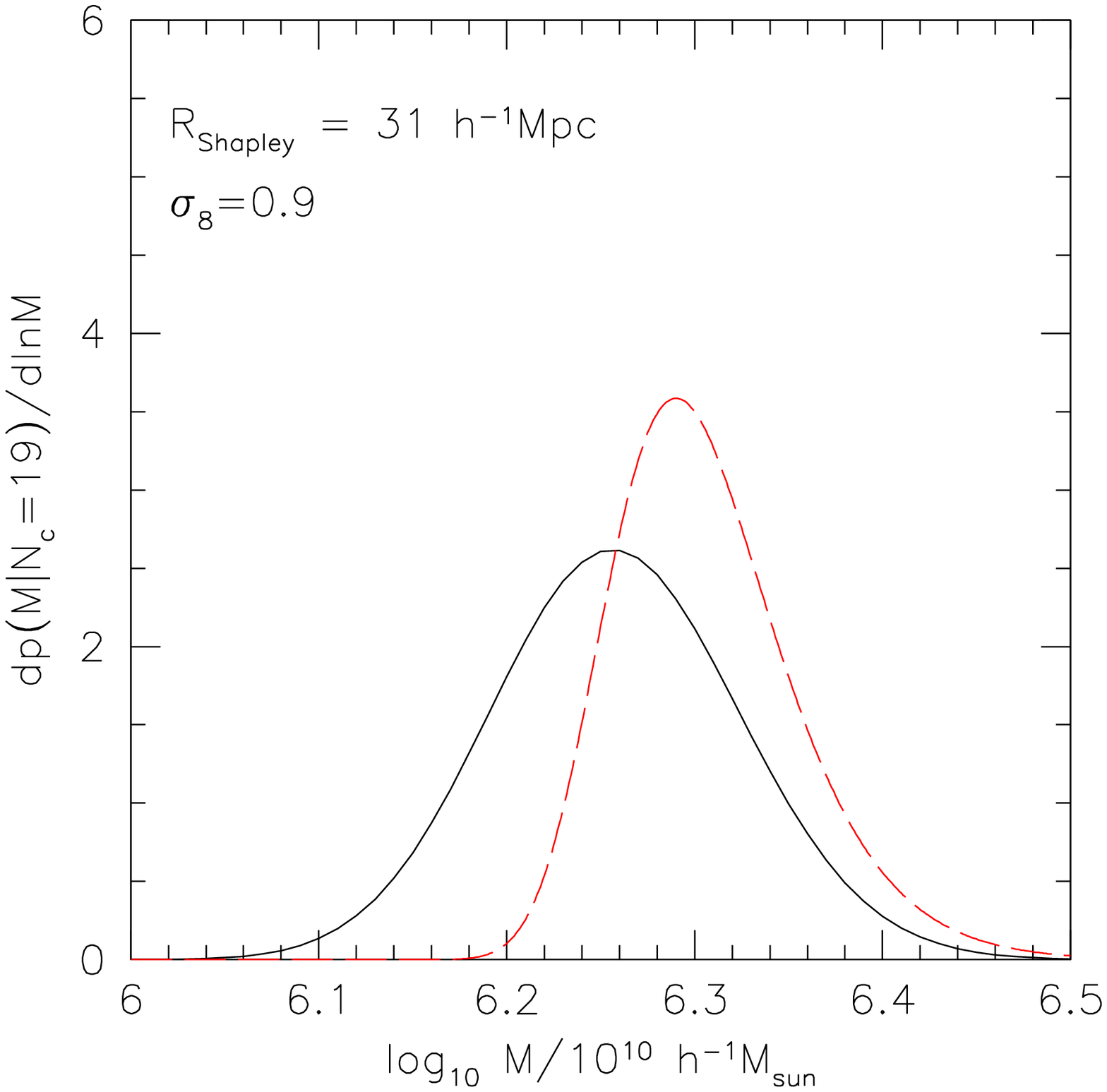}
 \caption{Dependence of the estimated mass of the Shapley supercluster 
          (solid) and of the densest of $(200/31)^3$ randomly placed 
           Shapley-sized cells (dashed) on $\sigma_8$.}
 \label{fig:shapley7-9}
 \end{center}
\end{figure}

Figure~\ref{fig:meanN} shows how $\bar N_\delta$, computed following 
Sheth \& Tormen (2002), increases with total mass $M$ for our three choices of 
$\sigma_8$.  This, in equation~(\ref{pMNV}), allows us to constrain 
the expected values of $M$.  The solid curve in 
Figure~\ref{fig:shapley8} shows $p(M|N,V)$ when $\sigma_8=0.8$.  
Figure~\ref{fig:shapley7-9} shows $p(M|N,V)$ for $\sigma_8=0.7$ (top) 
and $\sigma_8=0.9$ (bottom).  In effect, these are estimates 
of the total mass, and hence overdensity, of Shapley.  
Notice that these distributions shift slightly with $\sigma_8$.  
The sense of the trend is easily understood:  When $\sigma_8$ 
is small then massive halos are rare, so the environment must be 
that much more extreme to produce the observed number of clusters.  
At the peak values
 % $\log (M/h^{-1}M_\odot) = (16.26,16.24,16.22)$ 
 $\log (M/h^{-1}M_\odot) = (16.28,16.26,16.25)$ 
the associated overdensities are
 % $(1+\delta) = (1.98,1.89,1.81)$ 
 $(1+\delta) = (2.07,1.99,1.93)$ 
so the linear theory overdensities are
 $\delta_{\rm L}=(0.60,0.56,0.54)$, making 
 $(\delta_{\rm L}/\sigma_{\rm L})=(2.60,2.15,1.86)$.
These indicate that Shapley is not particularly unusual.\footnote{For 
$\sigma_8=0.8$, our estimate of $\delta_{\rm L}/\sigma_{\rm L}$ is 
close to that of Mu{\~n}oz \& Loeb (2008); our estimates of the total 
mass differ because they used a substantially larger volume estimate 
than do we.}   We argue in Section~\ref{sec:extremeIC} that to 
estimate the initial `peak height', it may be more appropriate to 
use $\sigma_{\rm L}(M)$ rather than $\sigma_{\rm L}(\bar\rho V)$.  
This yields higher values:  
$\delta_{\rm L}/\sigma_{\rm L} = (3.35,2.75,2.33)$. 
All these results are summarized in Table \ref{tab:SSC}.  
It is remarkable that our analytic estimate of the total mass 
is so similar to that derived by Ragone et al. (2006) using mock 
catalogs:  for $\sigma_8=0.9$ (the value in their mocks), 
our estimate is only 10\% larger than theirs.  

Upon evaluating an integral that is very similar to the one which 
defines $\bar N_\delta$, the excursion set approach also yields estimates 
of the typical mass fractions in such clusters.  If we use $f_\delta$ 
to denote this fraction, then 
\begin{equation}
 \bar f_\delta = \int_{M_{\rm min}}^M {\rm d}m \, N(m,\delta_c|M,V)\,(m/M).  
 \label{fdelta}
\end{equation}
At the peak values shown in the Figures,
 $\bar f_\delta =$ (0.14, 0.18, 0.22) for $\sigma_8=(0.7,0.8,0.9)$.  
Since the total observed mass in these 19 clusters is
 $5.27\times 10^{15}h^{-1}M_\odot$,
these mass fractions suggest total Shapley masses of
 $\log (M/h^{-1}M_\odot) = (16.57, 16.47, 16.38)$.  These values are
larger than the peak values from the excursion set approach, because 
the expression above assumes that the observed number of clusters is 
equal to $\bar N_\delta$, whereas it is actually larger by a factor of 
$(1.7,1.6,1.5)$.  Increasing $\bar f_\delta$ by these factors reduces 
the estimated total Shapley mass to 
 $\log (M/h^{-1}M_\odot) = (16.34, 16.26, 16.20)$.  
These values are in excellent agreement with our estimate above, 
which was based on the fact that 19 massive clusters were observed, 
but no other information about their masses was used, though the 
agreement is best for $\sigma_8=0.8$.  

\section{Extreme value statistics}\label{extremes}
It is interesting to compare the mass estimates derived above with 
the mass associated with the densest of $N$ randomly placed cells, 
where $N$ is the ratio of Shapley's volume to that in which it was 
found.  If the masses agree, then this would suggest that although 
Shapley is extreme, it is not unusually so.  Note that, despite the 
similarity, this is a different question from the one which is more 
often asked:  
Is the region containing Shapley the densest of its size in the 
entire sphere centered on our galaxy which contains Shapley?  

Given a total survey volume, the mass of the densest of $N$ cells 
placed randomly in this volume (i.e., large compared to the cells) -- 
which we will estimate below -- is certainly smaller than the mass 
associated with the question that is more usually asked.  This is 
because one might think of this densest region as a particularly carefully 
placed cell.  In particular, one would have to throw a large number of 
cells (compared to $N$) before one lands in just the right position to 
find this densest region.  We discuss the difference between these two 
extreme value estimates in Section~\ref{peaks}.  Of course, both 
require an assumption about the volume within which Shapley was found.  
We will assume that this is a sphere with radius $200h^{-1}$Mpc, and 
will discuss how our results depend on this choice shortly (e.g. 
following equation~\ref{eq:extremevalue}).

If $P_1(<M|V)$ denotes the probability that the most massive of the
 $N = (200/31)^3$ 
regions of volume 
 $V=V_{\rm Shapley}$ 
that are within $200h^{-1}$Mpc is less massive than $M$, then 
$P_1(<M|V)$ must equal the probability that each of the 
$N\approx 270$ cells is less massive than $M$.  Thus 
\begin{equation}
 P_1(<M|V) = \int_0^M {\rm d}M\, p_1(M|V) \approx p(<M|V)^N,
 \label{P1cum}
\end{equation}
and, by taking the derivative, 
\begin{equation}
 p_1(M|V) \approx N\,p(M|V)\,p(<M|V)^{N-1}.
 \label{p1MV}
\end{equation}
Appendix~\ref{independence} discusses this approximation further.  

Before we use this expression, notice that if $M_{1/2}$ denotes the 
median value of the expected mass, i.e., that at which
 $P_1(<M_{1/2}|V)=1/2$, 
then 
\begin{equation}
 -\frac{\ln(2)}{N} \approx \ln[1 - p(>M_{1/2}|V)] \approx -p(>M_{1/2}|V),
\label{eq:extremevalue}
\end{equation}
where we have assumed that $p(>M|V)\ll 1$ in the tail of the distribution.  
This shows that the mass returned by our approach is approximately the 
same as that given by setting $Np(>M|V) = 1$ (because $\ln 2$ is of 
order unity), which makes intuitive sense.  
It also illustrates that the mass estimate depends on $N$:  
If the large $M$ tail falls exponentially, then 
 $M_{1/2}\propto \ln(N/\ln(2))$.  I.e., the expected mass increases 
approximately as $\ln(N)$, so the dependence on $N$, and hence on 
our assumption that $V$ is the comoving volume within $200h^{-1}$Mpc, 
is weak.

This means that one can devise a test which asks if the survey volume 
which is required to make a certain mass object the densest of its type 
does indeed contain only one such object.  
Alternatively, if the survey volume is known but the mass is not, 
then the assumption that the object is the most massive actually 
yields an estimate of its mass.  
We will show shortly that Shapley passes either of these tests for 
currently acceptable values of $\sigma_8$.  

Finally, we note that the mass estimate can be rather precise.  
If we use $M_{0.84}$ to denote the value of the mass below which $84\%$ 
of the probability lies, namely the value at $+1\sigma$, then, for 
an exponentially falling distribution in $M$, 
 $M_{0.84} \propto \ln(N/\ln(1/0.84))$, so 
\begin{equation}
  \frac{M_{0.84}}{M_{1/2}} = 1 + \frac{\ln(\ln(2)/\ln(1/0.84))}{\ln(N/\ln(2))}
                         = 1 + \frac{1.38}{\ln(N/\ln(2))}\; .
 \label{precision}
\end{equation}
For $N=1000$ the fractional error on $M_{1/2}$ is 0.19, and it decreases 
as $\ln(N)$ increases.

\subsection{Extremes in the initial conditions}\label{sec:extremeIC}
To illustrate the approach, suppose that the pdf associated with scale 
$V$ is a Gaussian with variance $\sigma_{\rm L}$.  Then the extreme-value 
mass and survey volume are related, through 
equation~(\ref{eq:extremevalue}), by  
\begin{equation}
 {\rm erfc}\left(\frac{\delta_L}{\sigma_{\rm L}\sqrt{2}}\right)
 = \frac{2\ln(2)}{V_{\rm Survey}/V}, 
 \label{erfcL}
\end{equation}
where $\delta_{\rm L}$ is related to $M/V$ by equation~(\ref{scapprox}).  
The previous section argued that, if $\sigma_8=0.8$, then, for an 
object like Shapley,
 $\sigma_{\rm L}=0.26$ and $\delta_{\rm L}/\sigma_{\rm L} = 2.15$.  
These values in equation~(\ref{erfcL}) imply $V_{\rm Survey}/V \approx 44$. 
Since this is substantially smaller than 270, there should be at least 6
other Shapley-like objects within $200h^{-1}$Mpc of us.  This is unlikely.  
Alternatively, requiring $V_{\rm Survey}/V = 270$ means 
$\delta_{\rm L}/\sigma_{\rm L} = 2.8$.  For $\sigma_{\rm L}=0.26$, 
the associated nonlinear overdensity is $1+\delta = 2.6$ making 
the estimated mass $10^{16.42}h^{-1}M_\odot$.  This is about 0.18~dex 
larger than that from the excursion set approach, indicating that 
although Shapley is a rich concentration, it is not more extreme 
than one would expect on the basis of random statistics.  Therefore, 
it would not be unexpected to find an even more extreme object of 
its volume in the local universe.

One can improve on these estimates by noting that if one is using the 
linear pdf, then the appropriate smoothing scale is not $V$ but the 
associated initial scale $M/\bar\rho$, and $\sigma_{\rm L}$ should also 
be computed on the scale $M/\bar\rho$ rather than $V$ \cite[e.g.,][]{ls08}.  
Since $\sigma_{\rm L}$ is smaller than before, $\delta_{\rm L}/\sigma_{\rm L}$ 
will be larger, and we now require 
\begin{equation}
 {\rm erfc}\left(\frac{\delta_L}{\sigma_{\rm L}(M)\sqrt{2}}\right)
 = \frac{2\ln(2)}{V_{\rm Survey}/V(1+\delta)}.
 \label{betterL}
\end{equation}
The result is that $V_{\rm Survey}/V \approx 232(1+\delta) = 464$.  
Thus, Shapley is consistent with being the densest of $(200/31)^3$ cells, 
so we should not be surprised if we find another comparable or even more 
massive object in a survey that is only slightly deeper.  
Alternatively, if we set $V_{\rm Survey}/V = 270$, then 
equation~(\ref{betterL}) requires Shapley's mass to be 
$10^{16.245}h^{-1}M_\odot$, which is in good agreement with the excursion 
set analysis.

\begin{table}
\begin{tabular}{ccccc}
\hline
 & & & Excursion & Extremes \\
 $\sigma_8$ & $\delta_{\rm L}/\sigma_{\rm L}$ & $\delta_{\rm L}/\sigma_{\rm L}(M)$ 
            &      $\log_{10}Mh/M_\odot$ & $\log_{10}Mh/M_\odot$   \\
\hline
 0.7 & 2.60 & 3.35 & 16.28 & 16.22 \\
 0.8 & 2.15 & 2.75 & 16.26 & 16.25 \\
 0.9 & 1.86 & 2.33 & 16.25 & 16.29 \\
\hline
\end{tabular}
\caption{Estimated initial fluctuation height and mass of the 
         Shapley supercluster.  The values listed in columns~2 and~3 
         show that this large concentration of galaxies is not unlikely.}
\label{tab:SSC}
\end{table}

\subsection{Extremes in the nonlinear field}
It is interesting to contrast this treatment, which uses extreme value 
statistics of the initial pdf, with an analysis based on the nonlinear 
pdf.  In the previous section, we used the fact that the Lognormal 
distribution (equation~\ref{pLN}) is a reasonably accurate model.  
In this case, the distribution of $\ln(M)$ is Gaussian, so the previous 
analysis goes through except that now 
\begin{equation}
 {\rm erfc}\left(\frac{\ln(M/\bar\rho V) 
                  + \sigma_{\rm L}^2/2}{\sigma_{\rm L}\sqrt{2}}\right)
 = \frac{2\ln(2)}{V_{\rm Survey}/V}.  
\end{equation}
The associated estimate for $V_{\rm Survey}/V = N$ given the excursion 
set mass of $10^{16.26}h^{-1}M_\odot$ and $\sigma_{\rm L}=0.26$ is 270.  
The small differences compared to the previous estimates can be 
understood as deriving from the fact that the term in brackets in 
the erfc above effectively makes Shapley a fluctuation of height 
$2.79$ (for $\sigma_8=0.8$).  

In fact, the distribution of the expected mass is skewed.  
Hence, to provide a more direct comparison with the mass estimates 
from the previous section, which we also expressed as distributions, 
the dashed curves in Figures~\ref{fig:shapley8} and~\ref{fig:shapley7-9} 
show equation~(\ref{p1MV}) for the same Lognormal distributions of 
$p(M|V)$ that we used in the excursion set calculation.  
The overlap between the solid and dashed curves is remarkable, 
given how very different these two methods are.  E.g., for this 
calculation, the most probable mass $M$ decreases as $\sigma_8$ 
decreases (dashed curves in Figure~\ref{fig:shapley7-9}), because 
small values of $\sigma_8$ mean that large deviations from the mean 
value are rarer; this trend is opposite to that for the excursion set 
approach, where small values of $\sigma_8$ mean massive halos are 
rarer, so the total mass $M$ from which to obtain the observed number 
of massive halos must be larger.  
So it is interesting that the match between these two approaches is 
slightly better for $\sigma_8=0.8$ than for the other two cases.  
When $\sigma_8=0.8$, then Shapley is consistent with being the most 
massive of a random set of regions of volume $V_{\rm Shapley}$ 
in the local Universe; 
if $\sigma_8=0.9$, then Shapley lies at the low-end of the expected 
extreme-mass distribution; 
if $\sigma_8=0.7$, then it lies at the high-mass end.

These curves show that, if it is the most extreme object within 
$200h^{-1}$Mpc, then the existence of Shapley is easily accomodated in 
models with high $\sigma_8$; even $\sigma_8=0.7$ is not problematic.  
On the other hand, if $\sigma_8=0.9$, then, we will not have to increase 
the survey volume much before we see another object that is more 
extreme than Shapley.  
However, if $\sigma_8=0.7$, then Shapley should be the most extreme 
object even in a volume that is larger by a factor of 2.  
It happens that there is indeed a very large structure in the volume 
which lies just beyond Shapley.  The next section studies this structure 
in more detail.  

But before we do, it is worth noting that our extreme value mass 
estimate is rather precise:  the widths of the dashed curves in 
Figures~\ref{fig:shapley8} and~\ref{fig:shapley7-9} are typically 
less than 0.1~dex.  While this level of precision may be surprising, 
we note that its origin is understood:  setting $N=270$ in 
equation~(\ref{precision}) yields a fractional uncertainty 
of 0.23, which corresponds to $0.1$~dex.

\subsection{Peaks and extremes}\label{peaks}
So far, the extremes we have been considering are associated with the 
statistics of randomly placed cells.  However, we noted that we are 
often more interested in ascertaining whether or not a particular object 
is an extreme outlier -- since we have determined the location and size 
of the object a priori, treating it as a randomly placed cell is no longer 
appropriate.  At least for sufficiently overdense extremes, there is a 
relatively straightforward way to account for this difference.  
This is because sufficiently overdense objects in the nonlinear density 
field typically correspond to large fluctuations in the initial field:  
i.e., $\nu_{\rm L}\equiv \delta_{\rm L}/\sigma_{\rm L}\gg 1$.   
For such objects, it should be a good approximation to assume they 
formed from high peaks in the initial field (also see discussion in 
Colombi et al. 2011).  
The expected number density of peaks above some $\nu_t$ (which we would 
like to estimate) is related to the probability that a randomly placed 
cell lies above this same threshold as follows.  Typically, one can move 
the cell which defined the peak around a little bit without significantly 
changing the height of the fluctuation in it.  If we think of this as 
defining a volume around each peak, then 
\begin{equation}
 P(\ge\nu_{\rm t}) = \frac{{\rm erfc}(\nu_{\rm t}/\sqrt{2})}{2} 
                  = {\rm vol}(\ge\nu_{\rm t})\, n_{\rm pk}(\ge\nu_{\rm t}). 
\end{equation}
If the peak was associated with smoothing scale $R_M$, then this volume 
satisfies 
\begin{equation}
 {\rm vol}(>\nu_{\rm t}) = 
 \frac{(2\pi)^{3/2}R_M^3}{(\gamma R_M/R_*)^3\,(\nu_{\rm t}^3 - \nu_{\rm t})}
 \qquad {\rm as}\ \nu_{\rm t}\to\infty
\end{equation}
\cite{bbks}.  This shows that the volume scales approximately as 
$\nu_{\rm t}^{-3}$, with prefactors that can be understood as follows.  
The volume of a Gaussian smoothing filter is $(2\pi)^{3/2}\,R_f^3$, 
so the numerator is the moral equivalent of what we have been calling 
the volume of the randomly placed cell in the initial conditions:
  $V(1+\delta)$.  
This means that 
\begin{equation}
 \frac{V_{\rm Survey}}{V(1+\delta)}\,P(\ge\nu_{\rm t}) =  
 % \frac{{\rm vol}(\ge\nu_{\rm t})}{V(1+\delta)}\,
 %  V_{\rm Survey}\,n_{\rm pk}(\ge\nu_{\rm t}) = 
  \frac{n_{\rm pk}(\ge\nu_{\rm t})\, V_{\rm Survey}}
       {(\gamma R_M/R_*)^3\,(\nu_{\rm t}^3-\nu_{\rm t})}\, .
\end{equation}
If we now replace the requirement that 
 $[V_{\rm Survey}/V(1+\delta)]\,P(\ge\nu_{\rm t}) = 1$ 
with the requirement that 
 $n_{\rm pk}(\ge\nu_{\rm t})\, V_{\rm Survey} = 1$ 
(see equation \ref{eq:extremevalue} and below), 
then this means that we now want 
\begin{equation}
 \frac{V_{\rm Survey}}{V(1+\delta)}\,P(\ge\nu_{\rm t}) = 
 \frac{1}{(\gamma R_M/R_*)^3\,(\nu_{\rm t}^3-\nu_{\rm t})}.
\end{equation}
Comparison with equation~(\ref{betterL}) shows that the required 
$V_{\rm Survey}$ is reduced by a factor proportional to 
$(\nu_{\rm t}^3-\nu_{\rm t})$.  
For scale-free spectra, $(\gamma R_M/R_*)^3 = [(n+3)/6]^{3/2}$, 
and, for the large smoothing scales of interest here ($\sim 30h^{-1}$Mpc), 
we can think of a $\Lambda$CDM model as having $n$ between $0$ and 
$-1$.  This makes the required $V_{\rm Survey}$ smaller by a factor of 
approximately $2^{3/2}/\nu_{\rm t}^3$ or $3^{3/2}/\nu_{\rm t}^3$.  
Alternatively, if $V_{\rm Survey}/V$ is fixed, then the associated value 
of $\nu_{\rm t}$, and hence the associated mass estimate, will be larger 
than before.  Although the relation between the value from the peaks 
calculation and that for random cells depends on $\nu_{\rm t}$, 
at $\nu_{\rm t}\sim 5$ (the high peaks of most interest here), the peaks 
calculation returns approximately 1 plus the value from the random cells 
calculation. 

\begin{figure}
 \includegraphics[scale=0.4]{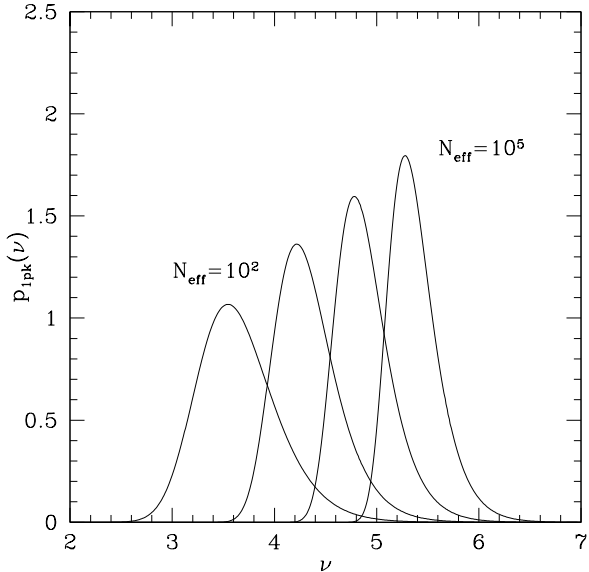}
 \caption{Dependence of the extreme value estimate of the height of 
          the highest peak on the ratio of survey to peak volumes.}
 \label{fig:expk}
\end{figure}

We can combine extreme value and peak statistics to make a slightly 
more detailed statement.  Namely, for a given ratio of survey to 
peak volume, what is the expected distribution of the height of the 
highest peak?  The same logic which led to equations~(\ref{P1cum}) 
and (\ref{p1MV}) implies that 
\begin{equation}
 p_{\rm 1pk}(\nu) \approx n_{\rm pk}(\nu)\,V_{\rm Survey}\,
                          {\rm exp}[-n_{\rm pk}(>\nu)V_{\rm Survey}].
 \label{p1pk}
\end{equation}
(The Appendix discusses how one might go beyond the Poisson/independent
cells assumption.)
Figure~\ref{fig:expk} shows this distribution for a number of choices of 
\begin{equation}
 N_{\rm eff}\equiv (\gamma R_{\rm pk}/R_*)^3\, 
                  \frac{V_{\rm Survey}}{(2\pi)^{3/2} R_{\rm pk}^3}.
\end{equation}
To make the plot, we have used the $\nu\gg 1$ approximation (4.14) of 
Bardeen et al. (1986) %\cite{bbks} 
rather than the full expression for $n_{\rm pk}(\nu)$, 
since we only expect this analysis to be valid for $\nu\gg 1$.  
But this does not affect the main point we wish to make:  that the 
height of the highest peak is only a weak function of $N_{\rm eff}$.  
This is the analogue of the statement we made previously about the 
weak dependence of $M_{1/2}$ on $N$.  The lesson is that very large 
survey volumes are required to reach large values of $\nu$.  

Note in particular, that this analysis is only valid for $\nu$ larger 
than the one given by the excursion set analysis of Shapley, so we 
will not make numerical estimates of these effects here.  However, in 
the next section, we will be interested in larger $\nu$, and this 
analysis will then be useful.  

\begin{figure*}
 \includegraphics[scale=0.95,angle=90]{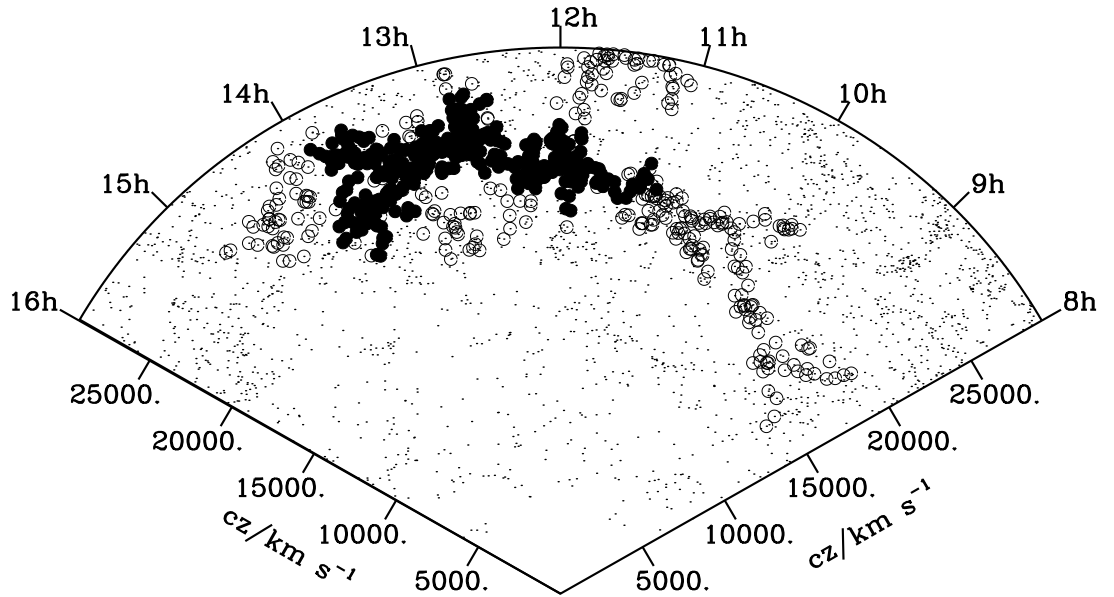}
 \vspace{-6cm}
 \caption{The Great Wall in the SDSS (filled circles), identified by 
          a percolation analysis of the SDSS percolation group 
          catalog (dots).  Open circles show the additional members 
          which are included if the percolation link length is 
          increased from $8h^{-1}$Mpc to $12h^{-1}$Mpc. }
 \label{fig:gw}
\end{figure*}

\section{The SDSS Great Wall}\label{sdss}
A dramatic structure at $z\sim 0.08$ is seen in the 2dF and SDSS 
galaxy surveys.  Now known as the Sloan Great Wall \citep{gott05}, 
it is, like Shapley, a region containing an overabundance of rich 
clusters.  We would like to perform a similar exercise to determine 
if it too can be easily accomodated in Gaussian theories.  
However, in this case, we do not yet have mass estimates of its 
members, and the appropriate lower limit in equation~(\ref{Ndelta}) 
is unknown. Therefore, we have extended our approach as follows.

\subsection{Percolation estimates of Wall volume}
We begin with the SDSS percolation catalog of groups in the SDSS 
\citep{sdss-groups}.  This provides a list of about $4100$ groups 
having three or more members brighter than $M_r=-19.9$.  We perform 
our own percolation analysis on this group catalog to identify the 
members of the Great Wall.  The size of the Wall depends on the 
parameters of our percolation analysis; we have found that a 
link-length of $8h^{-1}$~Mpc returns a catalog that closely 
corresponds to the contiguous structure picked out by eye.  
This is approximately given by
 $0.07\le z\le 0.092$ and $0\le {\rm dec}\le 6$ if $185\le {\rm ra}\le 210$
and 
 $0.07\le z\le 0.080$ and $0\le {\rm dec}\le 6$ if $166\le {\rm ra}\le 185$.
The underlying group catalog and the Great Wall members identified by 
our analysis are shown as dots and filled circles in Figure~\ref{fig:gw}.  
The Wall defined in this way contains 2180 galaxies in 335 groups.  
It has a volume of approximately $2.3\times 10^5 h^{-3}$Mpc$^3$, so 
its effective radius is about $38h^{-1}$Mpc; 
$\sigma_{\rm L} = 0.212\,(\sigma_8/0.8)$ on this scale.  

We note that the Wall appears to extend beyond the SDSS footprint 
towards negative declination.  Because this cut reduces our estimates of 
both the number of group members and the total volume, neither our 
excursion set nor our extreme value analyses are strongly affected by 
this cut.  

Our estimate of the total volume is determined from redshift-space 
quantities.  For a structure as large as this, the redshift-space 
volume is smaller than the real-space volume.  Figure~\ref{fig:gw} 
suggests that, along the line of sight, the structure varies from 
about 5000~km~s$^{-1}$ to about 2000~km~s$^{-1}$.  If we assume that 
line-of-sight velocities are unlikely to exceed 1000~km~s$^{-1}$, 
then the true structure may be larger in the redshift direction 
by a factor of between 1.2 and 1.5.  Hence, we may have underestimated 
the true volume of the Wall by this same factor.  In Section~\ref{exWall}, 
we will show that our conclusions about whether or not the Wall is 
unexpected are not very sensitive to this uncertainty. 

On the other hand, our choice of link-length makes the Wall 
significantly smaller in extent than claimed by Gott et al.  
Indeed, our estimate of the Wall's volume makes it only 
$(38/31)^3 = 1.8$ times larger than Shapley.  
A link-length of about $12h^{-1}$Mpc is required to get something 
approaching their definition (open circles).  In this case, the total 
volume is about $7.2\times 10^5 h^{-3}$Mpc$^3$ (effective radius 
$55h^{-1}$Mpc), $\sigma_{\rm L} = 0.139\,(\sigma_8/0.8)$, and the 
structure contains 3663 galaxies in 645 groups.  
Again, varying the total volume by $\sim 30\%$ makes little difference 
to the nature of our conclusions below.  More importantly, we will show 
that although our estimates of the mass in the Wall do depend strongly 
on the link-length used to define the Wall (the longer link-length yields 
a Wall with three times the volume, so one naively expects the mass 
to be about three times larger as well), our conclusions about how 
unusual the Wall is do not depend strongly on this choice.  

\subsection{A halo model-excursion set estimate of the Wall mass}
A halo model analysis of the underlying galaxy catalog (i.e., SDSS 
galaxies with $M_r<-19.9$) suggests that only halos above 
 $M_{\rm min} = 10^{12}M_\odot$ 
host such galaxies.  In halos of mass $m$ which host such galaxies, the 
probability of hosting $N_s$ additional galaxies (with $M_r<-19.9$) is 
given by a Poisson distribution with mean 
\begin{equation}
 \langle N_s|m\rangle = \left(\frac{m}{23\,M_{\rm min}}\right)^{1.16}
 \label{NsM}
\end{equation}
\citep{sdss-xig}.  To an excellent approximation, this relation 
between the galaxy population and halo mass is independent of 
environment \citep{as07}.  This is a key point, because it means 
that the relation above is expected to be as accurate for the halos 
in the Sloan Great Wall as elsewhere.  Moreover, this assumption has 
also been shown to accurately reproduce the properties of the galaxies 
in the percolation group catalog we are using here \citep{ssm07}.  

In the present context, the accuracy of the halo model decomposition, 
and of the Poisson distribution of $N_s$ in particular, 
means that we expect the fraction of halos of mass $m$ which host 
3 or more galaxies to be 
\begin{equation}
 f_3(m) = 1 - {\rm e}^{-\langle N_s|m\rangle}(1 + \langle N_s|m\rangle).
\end{equation}
Similar expressions for $f_n(m)$ can be defined for arbitrary $n$.  
Hence, the expected number of halos containing $n$ or more galaxies 
brighter than $M_r=-19.9$ that are in cells of volume $V$ containing 
total mass $M$ is 
\begin{equation}
 \bar N_\delta = \int_{M_{\rm min}}^M {\rm d}m \, N(m,\delta_c|M,V)\,f_n(m),
 \label{Ndgroups}
\end{equation}
where $M_{\rm min}=10^{12}h^{-1}M_\odot$ and $N(m,\delta_c|M,V)$ is 
the same quantity as before (c.f. equation~\ref{Ndelta}), but with 
the new value of $V$.  Indeed, the only significant difference from 
equation~(\ref{Ndelta}) is that we have now included a factor of 
$f_n(m)$ to account for the fact that only a fraction of halos of 
mass $m$ are expected to be in the group catalog.  Note that this 
factor does not depend on $M$ or $V$, because the large scale 
environment does not affect equation~(\ref{NsM}).  

With this expression for $\bar N_\delta$ in hand, we can now use 
equation~(\ref{pMNV}) along with the observed number $N$ of groups 
having $n$ or more galaxies, and our estimate of the total volume 
$V$ of the Great Wall to estimate its mass $M$.  For $\sigma_8=0.8$, 
the rms fluctuation on scale $V$ in linear theory is 
$\sigma_{\rm L}=0.142$.  As before (equation~\ref{poisNMV}), we 
assume a Poisson distribution for the number of groups, but now 
with mean given by equation~(\ref{Ndgroups}).\footnote{This follows 
from the Poisson assumption for halo counts in cells 
$(M,V)$, the fact that a random subsample of a Poisson distribution 
is Poisson, and because the distribution of the sum of Poisson 
distributed numbers is Poisson with mean given by the sum of the 
means of the individual distributions.}  An important check on 
our approach is to perform this analysis for a range of values of 
$n$:  the inferred mass distribution should not be sensitive to 
this choice.  
%For $n=(3, 4, 5, 6, 7, 8, 9, 10)$ the observed number 
%of groups is $N_{\rm groups}=(455, 266, 170, 125, 96, 83, 72, 60)$.
For $n=(3, 4, 5, 6, 7, 8, 9, 10)$ the observed number 
of groups is $N_{\rm groups}=(335, 199, 132, 96, 75, 68, 60, 49)$ 
when the link-length is $8h^{-1}$Mpc.  For the longer link-length $12h^{-1}$Mpc, 
$N_{\rm groups}=(645, 361, 219, 155, 117, 96, 84, 69)$.

\begin{figure}
 \begin{center}
 \includegraphics[scale=0.4]{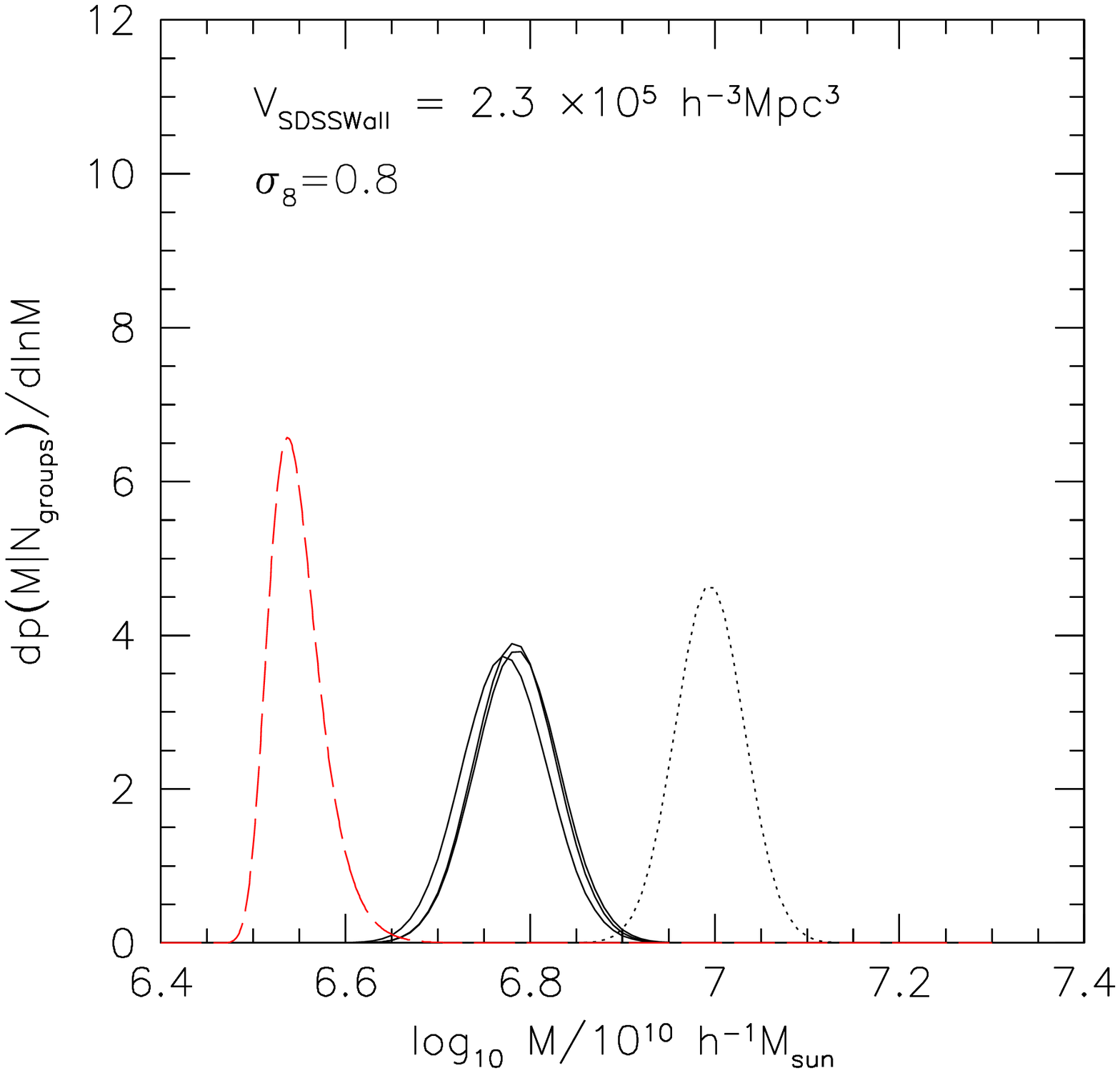}
 \includegraphics[scale=0.4]{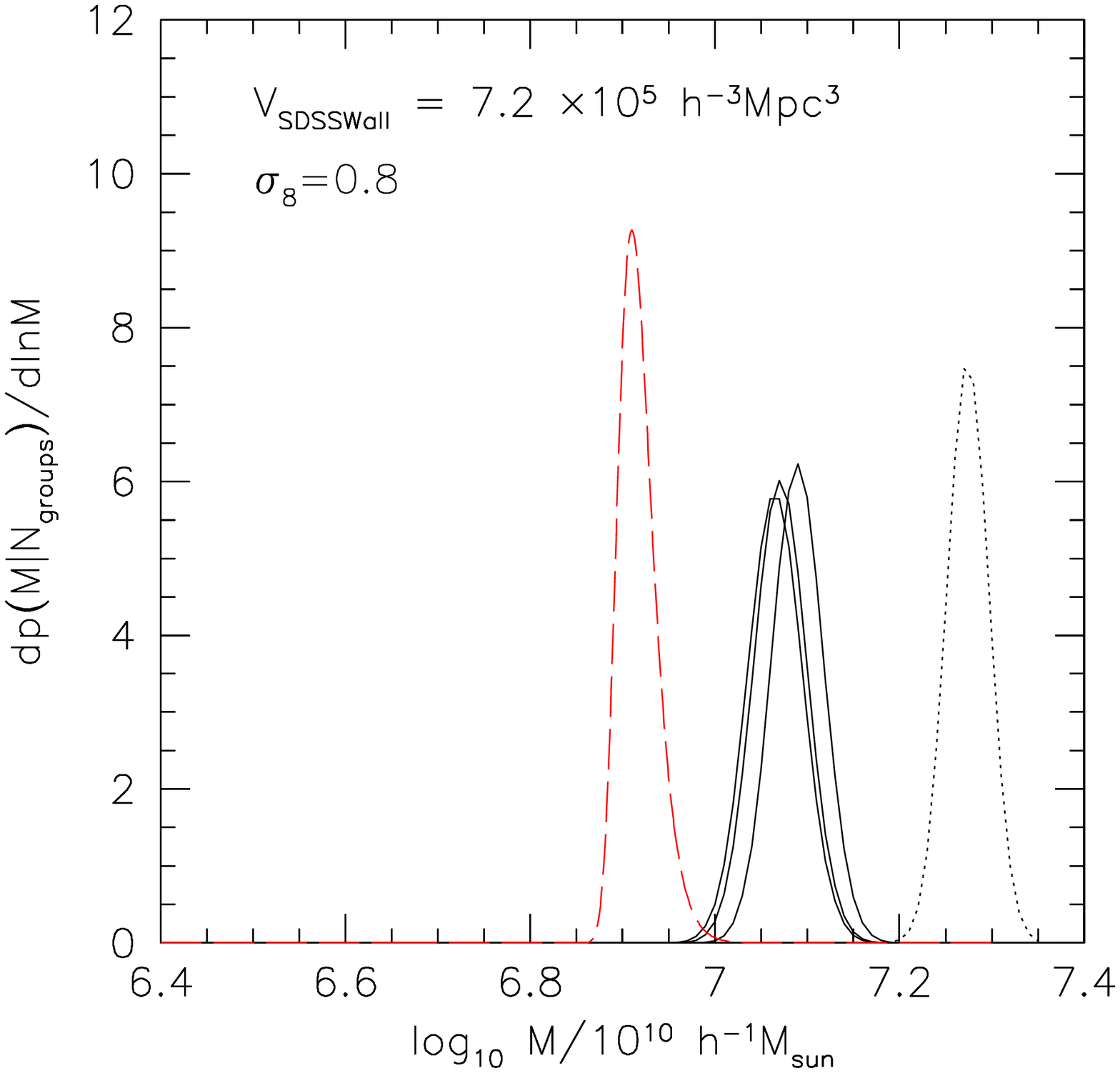}
 \caption{Comparison of the excursion set estimate of the mass of the 
          Sloan Great Wall (solid) with the expected mass returned  
          from the extreme value statistics approach (dashed)
          if $\sigma_8=0.8$.  
          Top and bottom panels show results when the Wall and its 
          members are defined using link-lengths of $8h^{-1}$Mpc and 
          $12h^{-1}$Mpc, respectively.
          Different solid curves in each panel show the excursion set 
          results for groups having more than $n=6,7$, and $8$ members;
          the dotted curve is for $n=3$.  
          The excursion set mass estimate shifts to lower masses as $n$  
          increases, although it is quite stable around $n=7$; 
          it is significantly larger than the estimate from extreme 
          value statistics.}
 \label{fig:sdsswall8}
 \end{center}
\end{figure}

The curves in the top panel of Figure~\ref{fig:sdsswall8} show a 
number of estimates of the mass of the Great Wall, when the 
link-length is $8h^{-1}$Mpc and $\sigma_8=0.8$.  The dotted curve, 
which is shifted towards larger masses than any of the other curves 
is for $n=3$.  This offset may be due to the difficulties associated 
with identifying small groups.  For $n>5$, the distributions overlap:  
we have shown $n=6,7$ and 8.  This is a nontrivial self-consistency 
test of our method.  However, at $n>10$ (not shown) the distributions 
shift further towards smaller masses; it may be that here we are in 
the regime of small number statistics, where the number of groups 
contributing to the estimate has dropped below 50, so that Poisson 
errors on $N_{\rm group}$ are more than 10\% of $N_{\rm group}$.  

\begin{figure}
 \begin{center}
 \includegraphics[scale=0.433]{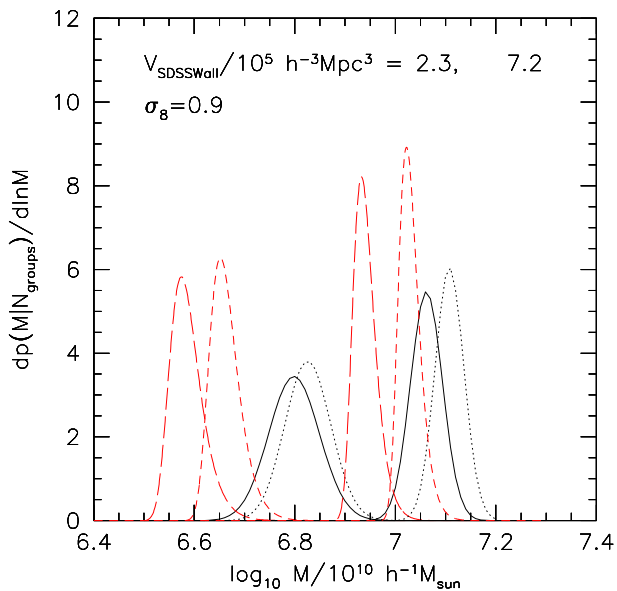}
 \caption{Similar to previous figure, but now $\sigma_8=0.9$.
          The left-most long dashed curve shows the 
          extreme value result for the shorter ($8h^{-1}$Mpc) link-length, 
          the short dashed curve just to the right of it shows 
          the result of increasing $V_{\rm Wall}$ by 30\%, to 
          approximately account for $z$-space effects.  
          The solid and dotted curves to the right of this curve 
          show the corresponding excursion set estimates (we 
          only show the $n=7$ result).  
          The next set of long- and short-dashed, solid, and 
          dotted curves show these same quantities when the Wall 
          is defined by the longer ($12h^{-1}$Mpc) link-length.}
 \label{fig:sdsswall9}
 \end{center}
\end{figure}

These curves suggest that the total mass in the Wall is about 
$10^{16.77}h^{-1}M_\odot$, meaning that the structure is about 3.55 
times denser than the background.  This in equation~(\ref{scapprox}) 
gives the associated linear theory density $\delta_{\rm L}$.  
In terms of the linear theory rms on this scale, we find 
 $\delta_{\rm L}/\sigma_{\rm L}=4.2$.  Using $\sigma_{\rm L}(M)$ 
instead makes this $6.6$.  
The overdensity in halos depends on $n$; 
it has 10 times the expected number of halos when $n=9$, 
but 9 times the expected mean number when $n=7$.  This is 
consistent with the fact that dense regions are expected to be 
overabundant in massive halos, and increasing $n$ removes lower 
mass halos.
The associated mass fraction in the observed groups (equation \ref{fdelta}) 
varies from about 40\% for $n=4$ to about 30\% for $n=9$.  

The corresponding results when the Wall is defined by the longer 
link-length are shown in the bottom panel.  
In this case, the total mass in the Wall 
is $M=10^{17.1}h^{-1}M_\odot$, so it is 2.25 times the background mass 
density, making $\delta_{\rm L}/\sigma_{\rm L}=4.6$; using 
$\sigma_{\rm L}(M)$ instead makes this $6.3$.  The overdensity in halos 
is about 5, and the observed groups account for about 20 percent of 
the total mass.  This smaller mass fraction is a direct consequence 
of defining the Wall as a looser structure.

\subsection{Extreme value statistics}\label{exWall}
The dashed curves in the two panels show the estimate of the mass 
associated with the extreme value statistics argument of 
Section~\ref{extremes}.  
This estimate requires as input the total survey volume, which 
we have set equal to the total comoving volume within $z=0.2$, 
making $V_{\rm Survey}/V_{\rm Wall} = 3456$ and $1100$ for the two 
(short and long) linking lengths.  
In contrast to when we performed this analysis for the Shapley 
supercluster, the dashed curve now lies to the left of the 
solid curves:  the excursion set estimates of the mass significantly 
exceed those expected based on extreme value statistics.  This 
means that, if the excursion set estimates are reliable, then the 
existence of the Wall is difficult to reconcile with the standard 
model.  

Increasing $\sigma_8$ alleviates the discrepancy slightly, 
as Figure~\ref{fig:sdsswall9} illustrates (solid and long-dashed curves).
If $\sigma_8=0.9$ and $n=7$, then the excursion set analysis of the 
structure defined by the $8h^{-1}$Mpc link length estimates a mass 
overdensity of $3.7$, a halo overdensity of $7.7$, 
$\delta_{\rm L}/\sigma_{\rm L}=3.8$ and 
$\delta_{\rm L}/\sigma_{\rm L}(M)=6.2$.  
These numbers are $2.2$, $4$, $4.04$ and $5.5$ when the link 
length is $12h^{-1}$Mpc (Table \ref{tab:SGW}).  For either structure, 
these are significantly larger than the extreme value estimate of the 
expected mass of the densest object.  

The second set of curves associated with each estimate (short-dashed and dotted lines) show the result 
of accounting crudely for redshift-space effects by increasing the 
Wall volume by 30\%.  To first order, increasing the volume increases 
all the mass estimates, but does not change the discrepancy between 
the extreme value and excursion set estimates.  This is the basis for 
our claim earlier that accounting for $z$-space distortions does not 
change our conclusions.  A more careful look shows that, the extreme 
value and excursion set mass estimates shift upwards by slightly 
different amounts:  about 0.1 and 0.05~dex, respectively.  
As a result, although the peaks are still quite well-separated, the 
tails of the mass estimates overlap slightly more.  
This means that the tension between excursion set and extreme value 
masses is alleviated somewhat, particularly for the $12h^{-1}$Mpc 
link-length.  

Thus, however we define it, the Wall is substantially more massive 
compared to the expected mass of the densest of $V_{\rm Survey}/V_{\rm Wall}$ 
randomly placed cells.  This can be appreciated directly from the fact 
that the excursion set analyses returned estimates of 
$\delta_{\rm L}/\sigma_{\rm L} \approx 4$ for the Wall, compared to 
$\approx 2$ for Shapley (for $\sigma_8=0.8$), even though 
$V_{\rm Survey}/V_{\rm Wall}$ is not much larger than $(200/31)^3$.  

It is interesting, therefore, to ask if its mass is also difficult 
to reconcile with the peaks model of Section~\ref{peaks}, which 
attempts to account for the fact that the Wall is not just a randomly 
placed cell.  In this case, an object with the mass and volume of 
the Wall would not be unusual only if it is the largest structure 
within a few times $10^8\, V_{\rm Wall}$; i.e., essentially within 
the Hubble volume.\footnote{We used $\delta_{\rm L}/\sigma_{\rm L}(M)\sim 6.5$ 
rather than $\delta_{\rm L}/\sigma_{\rm L}\sim 4$ to make this estimate.  
The Lognormal estimate of the effective peak height, $5.9$, is not 
very different.  Figure~\ref{fig:expk} shows that large $N_{\rm eff}$, 
and hence large volumes, are required to see even one peak of this 
height.}  Expressed another way, if $\sigma_8=0.8$ then the expected 
mass of the most extreme peak within $z=0.2$ is $10^{16.57}$ or 
$10^{16.95}$ for our two definitions of the Wall.
Although these are slightly larger than the randomly placed cells 
estimate, they are significantly smaller than the excursion set 
estimate.

\begin{table*}
\begin{tabular}{cccccccc}
\hline
 & & & & & & Excursion & Extremes \\
 $V/10^5 h^{-3}{\rm Mpc}^3$ & $\sigma_8$ & $\delta_{\rm L}/\sigma_{\rm L}$ &
 $\delta_{\rm L}/\sigma_{\rm L}(M)$ & $1+\delta_M$ & $1+\delta_n$ &
 $\log_{10}Mh/M_\odot$ & $\log_{10}Mh/M_\odot$   \\
\hline
 2.3 & 0.8 & 4.2  & 6.6 & 3.55 & 9 & 16.77 & 16.54  \\
 2.3 & 0.9 & 3.8  & 6.2 & 3.70 & 8 & 16.80 & 16.57  \\
 7.2 & 0.8 & 4.6  & 6.3 & 2.25 & 5 & 17.07 & 16.91 \\
 7.2 & 0.9 & 4.04 & 5.5 & 2.20 & 4 & 17.07 & 16.94 \\
\hline
\end{tabular}
\caption{Estimated initial fluctuation height, mass overdensity, galaxy 
         overdensity and mass of the SDSS Great Wall.  The two upper rows
         refer to the $8h^{-1}$Mpc link length; the two lower rows to 
         $12h^{-1}$Mpc. }
\label{tab:SGW}
\end{table*}

\section{Discussion and an extension}
We discussed a number of methods for estimating the masses of 
extreme objects in the Universe, and applied them to two of the 
most dramatic objects in the local Universe:  the Shapley 
supercluster and the Sloan Great Wall.  We used a percolation 
analysis to define these systems, and illustrated how our results 
depended on the link-length ($8$ or $12h^{-1}$Mpc) used to 
define it.  

In the case of Shapley, our estimate of the mass comes from combining 
estimates of the masses of its constituents with an excursion 
set analysis of the depedence of the halo mass function on the density 
of the local environment.  Unfortunately, this was not possible in the 
case of the Wall, since mass estimates of its constituents are not 
available.  In this case, we combined the excursion set analysis with 
a Halo-Model interpretation of its constituent groups, themselves 
identified from (optical) SDSS redshift survey data.  
Unfortunately, this method cannot currently be applied to Shapley, 
since it lies outside the SDSS footprint.  This is also why we have 
not included results from the recent analyses of the Wall by 
Einasto et al. (2010, 2011) -- but we hope to do so soon.  

We compared these mass estimates with that expected for the densest 
object in an appropriately defined `local' universe, and argued that 
the existence of Shapley is easily explained by currently popular 
models of structure formation (Figures~\ref{fig:shapley8} 
and~\ref{fig:shapley7-9}); its mass ($1.82\times 10^{16}h^{-1}M_\odot$) 
is consistent with it being the most massive object of its volume 
($1.25\times 10^5h^{-3}$Mpc$^3$) within $200h^{-1}$Mpc.  
% Our completely analytic mass estimate is within ten percent of that 
% made by Ragone et al. (2006) on the basis incompleteness corrections
% calibrated using mock catalogs.  

On the other hand, the Sloan Great Wall (Figure~\ref{fig:gw}) is 
difficult to explain, especially if the amplitude of the initial 
fluctuation field was at the low end of currently accepted values 
(Figures~\ref{fig:sdsswall8} and~\ref{fig:sdsswall9}).  Its mass
 $(5.9, 12.6)\times 10^{16}h^{-1}M_\odot$ 
is larger than expected for the most massive object of its volume
 $(2.3, 7.2)\times 10^{5}h^{-3}$Mpc$^3$ within $z=0.2$ 
(where the two numbers are for defining the Wall using 
link-lengths of $8$ or $12h^{-1}$Mpc respectively).   
If $\sigma_8=0.8$, then insertion of the excursion set estimate of 
its mass in our extreme value statistics calculation suggests that 
it must be the densest object of its volume within the Hubble 
volume.  An analysis which combines the excursion set estimate of 
the initial overdensity associated with the Wall, 
$\delta/\sigma\approx 6$, with the assumption that this fluctuation 
was the largest peak in the initial conditions, leads to a similar 
conclusion (Figure~\ref{fig:expk}).  

We are hesitant to make strong statements about whether this makes 
the Great Wall inconsistent with Gaussian initial conditions with 
acceptable values of $\sigma_8$, primarily because our current numbers 
are based on assuming the Wall is spherically symmetric when it clearly 
is not.  For this reason, we are in the process of extending both our 
methods -- the excursion set and extreme value statistics analyses -- 
to account for this.  Here we are aided by the fact that the Wall 
itself is not virialized.  Hence, we can use the simple parametrization 
of triaxial collapse from Lam \& Sheth (2008) to generalize 
equation~(\ref{scapprox}) for the mapping between nonlinear and linear 
overdensity.  This can then be used in our excursion set analysis.  
With this estimate of initial overdensity and shape in hand, we can 
modify our extreme value statistics calculation by replacing the number 
density of initial density of peaks of specified scale and height by 
adding the constraint that comes from specifying the shape 
\cite[e.g.,][]{bbks}.  This is the subject of work in progress.  

Our results suggest that the Sloan Great Wall is about 5 times the 
volume and about the same factor times the mass of the Shapley 
supercluster (we have used the larger mass and volume estimates of 
the Wall).  So one might wonder if Shapley is about the sixth most 
extreme object of its volume within $z=0.2$.  
It is straightforward to extend our application of extreme value 
statistics to address this question.  In particular, the same logic 
which leads to equation~(\ref{p1MV}) implies that the expected 
distribution of the mass of the $n$th densest region is 
\begin{eqnarray}
 p_n(M|V) &\approx& {N\choose n}\,n\,p(M|V)\,\Bigl[1-p(<M|V)\Bigr]^{n-1}
            \nonumber\\
          &&\qquad\qquad \times \quad p(<M|V)^{N-n}
\end{eqnarray}
\citep[e.g.][]{gumbel}.
The dotted curve in Figure~\ref{fig:shapley8} shows this 
expression, evaluated with $n=6$, $N=6375$, and $\sigma_{\rm L}=0.24$.  
This shows that Shapley could easily be the sixth most massive object 
within $z=0.2$ if $\sigma_8 = 0.8$.  Of course, it is trivial to 
extend this to our extreme value treatment of peaks:  one simply 
replaces $p(<M|V)\to \exp(-n_{\rm pk}(>\nu)\,V_{\rm Survey})$.  
The Appendix discusses how to modify this approach to account for 
the clustering of peaks.

Similarly, one can write down expressions for the joint probability 
distribution of the masses of e.g., Shapley and the Great Wall, if 
we require one to be the $i$th and the other the $j$th most extreme 
object of its type (recall they may have different values of 
$\sigma_{\rm L}$) in the same survey volume -- although we have not 
reproduced them here.  

One of the surprises of these analyses is, perhaps, the precision 
of the mass estimates it returns:  typically, these are of order 
15\%, both for the excursion set and the extreme value statistics 
approaches.  Although we provided some analysis for why this is 
so (equation~\ref{precision}), it would have been nice to test our 
mass estimates by combining the motions of the clusters in these 
systems with an infall model.  However, because the Shapley 
supercluster and the Sloan Great Wall are both far from round 
(e.g. Section~\ref{Shapley}), estimates based on the spherical 
collapse model are inappropriate.  Therefore, we are currently in 
the process of developing an infall model based on the assumption 
of a triaxial collapse.  

The precision of the mass estimates derives from the fact that the 
extreme fluctuations we are considering are from Gaussian random fields, 
in which extreme fluctuations are rare, so the distribution of events on 
the tail will be similar to one another.  However, it is almost certain 
that, at least for the extreme value statistics calculation, this is 
more generic.  This is because a large class of initial distributions 
have, as their limiting extreme value statistic, a double-exponential 
form \citep{ft28, gumbel}.  In the astrophysical context, this 
Fisher-Tippet or Gumbel distribution, and the study of extreme 
value statistics in general, has a long history in the study of 
the brightest galaxies in clusters \citep{scott, bb85}.  
Our work suggests that extreme value statistics may continue to 
provide insight into the study of the largest structures in the Universe.  

In particular, it would be interesting to use this approach to see if the 
sizes of the largest voids, or the masses of the most massive clusters 
or superclusters (e.g. Luparello et al. 2011; Schirmer et al. 2011; 
Yaryura et al. 2011), are consistent with the hypothesis that the 
initial fluctuation field was Gaussian.  
To use our approach for more generic initial conditions, one must know 
how the halo mass function depends on the large scale environment and 
one must have a model for the nonlinear probability distribution 
function.  For non-Gaussian initial conditions of the local type, 
such models have recently become available \citep{ls09}.

\section*{Acknowledgements}
We thank the INFN exchange program for support, 
and the Centro di Ciencias de Benasque ``Pedro Pascual'' for hospitality 
during the summer of 2008 when most of this work was completed, 
J{\"o}rg Colberg for encouragement then, 
Aseem Paranjape for encouragement now, and 
Stefano Camera for providing the opportunity for us to meet again 
and complete this work.  RKS is supported in part by NSF-AST 0908241. 
AD acknowledges additional support from the INFN grant PD51 and the 
PRIN-MIUR-2008 grant ``Matter-antimatter asymmetry,
dark matter and dark energy in the LHC era''. 
This research has made use of NASA's Astrophysics Data System.

\appendix

\section{On the approximation of independent cells when calculating 
         extreme values of spatial statistics}\label{independence}

The calculation of extreme value statistics reduces to one of writing 
the probability that, of $n$ draws from a distribution, none are above 
a certain value.  This raises the question of whether or not the draws 
can be assumed to be independent picks.  For the spatial statistics we 
are considering here, in which each cell represents a pick, and the 
total volume is the sum of the cells, the answer is clearly `no' because 
there are correlations between the cells.  
On the other hand, since the correlations decrease with cell separation, 
most cells will only be strongly correlated with a few nearby cells.  
Moreover, since we will generally be interested in large cells, even 
nearby cells are likely to be only weakly correlated.  
So the assumption of independence, may in fact be quite good.  
The question is:  Are extreme value statistics likely to be distorted 
by even these weak correlations?  After all, the whole point of such 
stastistics is that they are sensitive to the tails of the distribution, 
and these are where (fractional) changes to the distribution will be 
largest.  In what follows, we quantify this effect.  

To proceed, we need an expression for the joint distribution of 
$n$-draws.  We will first use a multivariate Gaussian to illustrate 
the argument, and then discuss possible generalizations.  If $\delta_i$ 
denotes the value of the field at position $i$, then the multivariate 
Gaussian distribution is specified by the covariance matrix ${\bm C}$, 
the elements of which are $C_{ij}=\langle\delta_i\delta_j\rangle$ (we 
are assuming $\langle\delta_i\rangle=0$ for all $i$).  In our case, 
$C_{ij}$ will be a function of the separation $r$ between cells $i$ 
and $j$.  Namely, 
\begin{eqnarray}
  C_{ij} &\equiv& \frac{\sigma_{ij}^2(r)}{\sigma_{ii}(0)\sigma_{jj}(0)} 
                 \qquad{\rm where} \\
  \sigma^2_{ij}(r) &\equiv&
          \int \frac{{\rm d}k}{k}\,\frac{k^3P(k)}{2\pi^2}\,W(kR_i)W(kR_j)\,
               \frac{\sin(kr)}{kr},\nonumber
\end{eqnarray}
and we have allowed for the fact that the cells of interest at position 
$i$ may have a different size than those at position $j$.  In the main 
text we were primarily interested in the case $R_i=R_j$.  If the $R_i$ 
are large, and/or the separation between cells is large, then ${\bm C}$ 
will be close to diagonal, so the $n$-point distribution will be 
well-approximated by the product of $n$ 1-point distribution functions.  
As a result, 
\begin{equation}
 \int_{-\infty}^{\delta_c} {\rm d}\delta_1 \cdots \int_{-\infty}^{\delta_c} 
    {\rm d}\delta_n \, p(\delta_1,\cdots,\delta_n) 
 \approx \prod_i^n \int_{-\infty}^{\delta_c}{\rm d}\delta_i \, p(\delta_i) .
 \label{joint2indep}
\end{equation}
This is the approximation used in equation~(\ref{P1cum}) of the main 
text.  The leading order correction to this can be obtained by 
writing this in terms of integrals above $\delta_c$, and then using 
previous results for high peaks or dense patches \cite{bbks,js86} 
to evaluate the result, which shows that the expression gets a correction 
factor which, to lowest order, depends on the two-point correlation 
function of regions above $\delta_c$.

In practice, the present day 1-point distribution function is no longer 
Gaussian.  However, on large scales, it may be a good approximation to 
assume that there is a monotonic mapping between the nonlinear overdensity 
and the linear one.  E.g., the main text assumes that this mapping is 
well approximated by a lognormal.  If one assumes that this is also true 
of the $n$-point distribution function, then we have a fully specified 
model of the nonlinear $n$-point function, expressed in terms of the 
initial Gaussian covariance matrix.  Now, the extreme value statistics 
care about the cumulative distribution:  the monotonicity of the mapping 
means that the net effect of nonlinear evolution is simply to shift the 
threshold of the corresponding multivariate (linear theory) Gaussian.  
Once this shift has been applied, then the previous analysis of the 
Gaussian case goes through in its entirety.  This justifies our use of 
equation~(\ref{P1cum}), and also shows how it might be improved.  

\subsection{Including the clustering of extrema}
Equation~(\ref{p1pk}) in the main text follows from the assumption that 
peaks are uncorrelated, so the probability that there are no peaks in 
$V_{\rm survey}$ is given by the Poisson expression
 $\exp(-n_{\rm pk}V_{\rm Survey})$. 
This can be derived from equation~(\ref{joint2indep}), by taking the 
limit of infinite sampling (in which $n\to\infty$, so the typical spacing 
between the cells is no longer of order their size).  
Going beyond the Poisson model requires a calculation of the higher 
order correlation functions (White 1979).  These are only known 
approximately (Appendix~F in Bardeen et al. 1986).  
On large scales where these are small, the required replacement 
in equation~(\ref{p1pk}) is 
\begin{displaymath}
 n_{\rm pk}(\ge\nu)V_{\rm Survey}\to n_{\rm pk}(\ge\nu)V_{\rm Survey} - 
              [n_{\rm pk}(\ge\nu)V_{\rm Survey}]^2 \frac{\bar\xi_{\rm pk}}{2},
\end{displaymath} 
where, for high peaks on large scales, 
\begin{eqnarray} 
 && [n_{\rm pk}(\ge\nu)\,V_{\rm Survey}]^2 \bar\xi_{\rm pk} \approx
  [n_{\rm pk}(\ge\nu)b_{\rm pk}(\ge\nu)V_{\rm Survey}]^2 \bar\xi \nonumber\\
   &\approx& \left[N_{\rm eff}\, (\nu^4 + \nu^2 + 2)\, 
           \frac{\exp(-\nu^2/2)}{\sqrt{2\pi}}\right]^2 \,
           \frac{\sigma_0^2(R_{\rm Survey})}{\sigma_0^2(R_{\rm pk})}. 
 %  &\approx& \left[N_{\rm eff}\, (\nu^4 + \nu^2 + 2)\, 
 %          \frac{\exp(-\nu^2/2)}{\sqrt{2\pi}}\right]^2 \,
 %          \left(\frac{R_{\rm pk}}{R_{\rm Survey}}\right)^{n+3}
\end{eqnarray}
Including this extra term affects the distributions shown in 
Figure~\ref{fig:expk} for $N_{\rm eff}<10^3$ or so (the peak shifts 
to slightly larger $\nu$) but matters little for larger $N_{\rm eff}$.

\label{lastpage}
\end{document}